# Interaction between substrate and probe in liquid metal Ga: Experimental and theoretical analysis

**Authors:** Ken-ichi Amano[1]✉, Kentaro Tozawa[1], Maho Tomita[1], Hiroshi Nakano[2], Makoto Murata[3], Yousuke Abe[3], Toru Utsunomiya[3], Hiroyuki Sugimura[3], and Takashi Ichii[3]✉

[1]Faculty of Agriculture, Meijo University, Nagoya, Aichi 468-8502, Japan

[2]Department of Molecular Engineering, Kyoto University, Kyoto 615-8510, Japan

[3]Department of Materials Science and Engineering, Kyoto University, Kyoto 606-8501, Japan

✉e-mail: amanok@meijo-u.ac.jp; ichii.takashi.2m@kyoto-u.ac.jp

**ABSTRACT:** Understanding the interaction between two bodies in a liquid metal is important for developing metals with high stiffness, strength, plasticity, and thermal stability. We conducted atomic force microscopy measurements in liquid Ga and performed a theoretical calculation in which the statistical mechanics of a simple liquid containing a quantum effect was used. The experiment and theory showed unusual behaviours in the interactions between the probe and substrate in the liquid metal. In the interactions, there were relatively numerous oscillations and large amplitudes. Furthermore, the interaction ranges were relatively long. From the theoretical calculations, we found an asymmetric property that when the probe is solvophilic and the substrate is solvophobic, the interaction tends to be repulsive; when the solvation affinities are exchanged, the interaction tends to be attractive in the close position. Our findings will be useful for understanding and controlling dispersion stabilities of nanoparticles and chemical reactions in liquid metals.

Solid metals have high stiffness, strength, plasticity, and thermal and electrical conductivities. Improvements in these properties are essential for the development of improved metal-based products. For example, high stiffness, strength, and plasticity are useful in artificial joints, reinforcing bars, and car bodies. In certain fields, nanoparticles are dispersed in solid metals to enhance the properties of these products. Hence, the dispersion stability of nanoparticles in liquid metals is an important topic[1]. Liquid metals are also used in batteries to avoid the growth of dendrites on electrodes[2] and used in piezoelectric elements to fabricate anisotropic piezoconductivity.[3] In addition, liquid metals are used in the production of thermal conductors (heat pipes) owing to their fluidity and high thermal conductivity. A heat pipe is applied to prevent permafrost melting[4] in nuclear reactors[4,5] and space reactor power systems [5]. In the liquid metals, there are problems of the aggregation and precipitation of impurities. From the viewpoint of these problems, the interactions between two bodies of liquid metals must be understood.

In previous studies, the effective pair potentials between constituent atoms in liquid metals have been studied theoretically[6,7] and experimentally[8,9,10]. A theoretical study[6,7] found that a characteristic oscillation exists in the effective pair potential. Similarly, the characteristic oscillation has been experimentally determined using small-angle X-ray scattering (SAXS) [8,9,10]. Interactions between *non*-constituent atoms in liquid metals have also been studied. For example, the potential of mean force (PMF) between the SiC nanoparticles has been thermodynamically predicted[1], and the total pair correlation function, which is used for calculation of the PMF, between the metallic solutes has been calculated using quantum mechanics simulations[11]. The density profiles of a liquid metal near a substrate have also been calculated using quantum mechanics simulation[12]. The thermodynamic prediction[1] mentioned above was performed in a simple and targeted manner. In the prediction, the PMFs at the contact point and at two slightly separated points were estimated. Starting from zero PMF at a sufficiently separated point, the curve of the PMF was drawn with a plausible smooth fitting[1]. In quantum mechanical simulations[11], ab initio molecular dynamics (AIMD) simulations were performed to obtain the total pair correlation function between the metal solutes in sodium liquid. The AIMD simulation is more theoretically accurate than the thermodynamic prediction. However, the system size, simulation time, and number of atoms were not sufficiently large owing to their high computational costs. As explained above, the thermodynamic and the quantum mechanical calculations have both merits and demerits. Therefore, we measured the interaction curves (force and PMF curves) between a probe and a substrate in liquid Ga using atomic force microscopy (AFM)[13] to investigate the true shape of the PMF (see Methods subsection 'AFM experiment'). To the best of our knowledge, this is the first time that interaction curves have been measured using AFM. Moreover, we theoretically studied the interaction using statistical mechanics of simple liquids[14,15,16,17] (see Fig. 1 and Methods subsections 'Pair potentials' and 'Integral equation theory'). In the calculation, the quantum effect in the bulk liquid Ga measured by SAXS[8] was incorporated. The system size and the number of atoms in the calculation were sufficiently large, unlike those in the AIMD simulation. In addition, the PMF curve calculated from the statistical theory was sufficiently continuous, unlike the thermodynamic prediction.

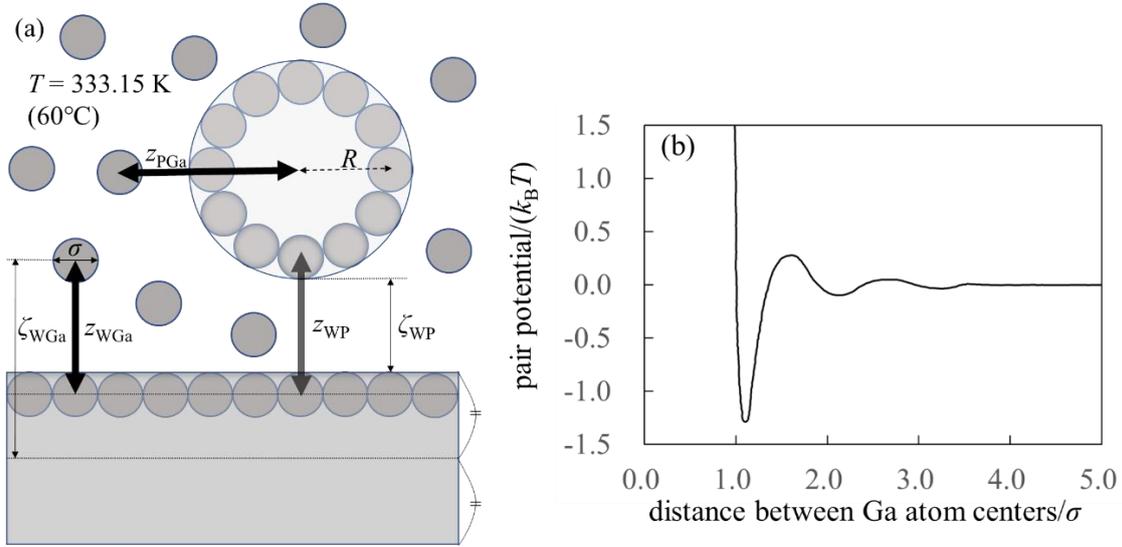

**Fig. 1| Pair potentials. a**, Schematic of the calculation system. Although the thickness of the substrate appears to be thin in the figure, it is 10 times the diameter of the Ga atom. **b**, The pair potential between Ga atom centres is the effective potential between the Ga cations in the sea of conduction electrons[8]. The diameter of the Ga atom is represented by $\sigma$ (= 0.255 nm)[8], and $k_B$ and $T$ are the Boltzmann constant and absolute temperature, respectively, where $T$ is 333.15 K (*i.e.* 60°C).

We obtained the force curve between the substrate and probe in liquid Ga using AFM (Fig. 2a). To the best of our knowledge, this is the first time the force curve was measured in a liquid metal. Mica and silica were used as the substrate and probe, respectively. However, mica and silica surfaces are coated with gallium oxide[13]. Hence, Fig. 2a represents the force curve between the gallium oxide surfaces[18,19] in liquid Ga (see Methods subsection 'AFM experiment'). The figure depicts relatively numerous oscillations (six to seven observable oscillations) and strong attractive forces. For example, three to four oscillations were observed in the general force curves[14,20]. The oscillation length was approximately equal to the effective diameter of the Ga atom. By integrating the force curve, we obtained a PMF curve (Fig. 2b). A relatively deep PMF minimum was observed in the vicinity of the substrate surface. The reason for the relatively strong attractive interactions in the experimental results (Fig. 2) is examined through the following theoretical calculations. For reference, we have included frequency shift curves (original data in Fig. 2) in the *Supplementary Information*. In addition, frequency shift, force, and PMF curves between the gallium oxide-coated probe and $AuGa_2$ surface in liquid Ga are also shown in the *Supplementary Information*.

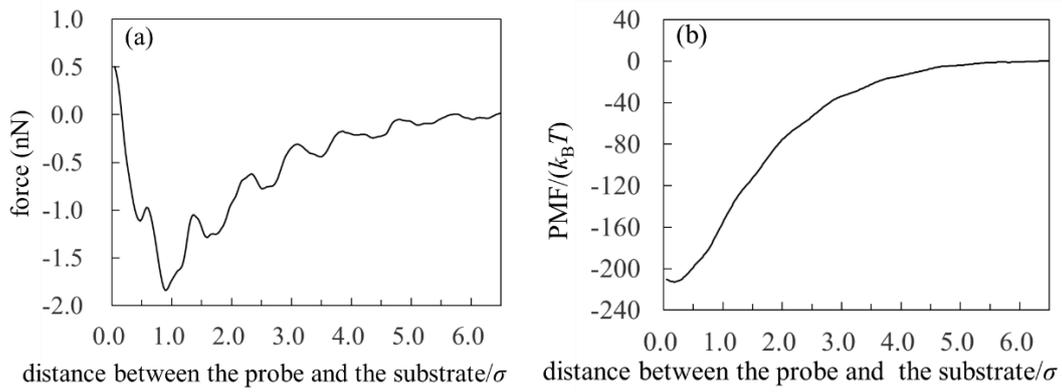

**Fig. 2| AFM results. a** and **b,** Force and PMF curves between the substrate and probe in liquid Ga, respectively, which are experimental results of the AFM.

We theoretically calculated the force curves, PMF curves, and density distributions of the Ga atoms near the substrate and probe using the affinity parameters $\varepsilon_{WGa}$ (= $10^{-22}$ or $75 \times 10^{-22}$ J) and $\varepsilon_{PGa}$ (= $10^{-22}$ or $75 \times 10^{-22}$ J). We defined values of the affinity parameters $10^{-22}$ J and $75 \times 10^{-22}$ J as "solvophobic" and "solvophilic", respectively. The solvophobic value was set so that minimum of the pair potential is almost nothing. The solvophilic parameter was set to the maximum value that can be handled by our integral equation theory. In the calculation, the diameter of the probe was five times the diameter of the Ga atom (*i.e.* $(2R + \sigma)/\sigma = 5$). The results when the diameter of the probe was 10 times the diameter of the Ga atom are shown in the *Supplementary Information*.

Fig. 3a (upper left) demonstrates the force curve between the solvophobic substrate and solvophobic probe. A relatively strong attractive force exists because the substrate and probe are both solvophobic. In other words, a solvophobic attractive interaction also exists in liquid Ga. A comparison of Fig. 2a with Fig. 3a (upper left) showed that they are qualitatively similar. For example, strong attractive forces and numerous oscillations were found to occur. The amplitudes of both force curves were also similar. We consider that the surfaces of the substrate and probe in the experiment are solvophobic[21] because they are coated with the gallium oxide[18,19]. This was attributed to the attractive solvophobic force.

Fig. 3a (bottom right) shows the force curve between the solvophilic substrate and solvophilic probe. As shown, the force curve also contains many oscillations, similar to other force curves. Unlike the force curve in Fig. 3a (upper left), there is a relatively strong repulsive force. This trend can be attributed to the surface affinities of the substrate and probe. Because both surfaces are solvophilic and want to solvate with liquid Ga as much as possible, they are stable when they are separated rather than in contact.

Fig. 3a (upper right) shows the force curve between the solvophobic substrate and solvophilic probe. The force curve also exhibits many oscillations, and its shape is similar to that shown in Fig. 3a (bottom right). However, the repulsive force in the force curve is weaker than that in Fig. 3a (bottom right). Fig. 3a (bottom left) shows the force curve between the solvophilic substrate and solvophobic probe, which also exhibits many oscillations. The repulsive and attractive forces were observed in the force curve. By comparing the force curves presented in the upper right and the bottom left, it is found that the former contains only repulsive force while the latter contains both repulsive and attractive forces. That is, the results are different despite mere exchange of the surface affinities of the substrate and probe. The results shown in Fig. 3a are qualitatively similar to those measured in an aqueous solution using AFM[20].

Fig. 3b (upper left) demonstrates the PMF curve between the solvophobic substrate and solvophobic probe. There is a negative attractive potential because both the substrate and probe are coated with the gallium oxide[18,19] (i.e. they are solvophobic surfaces[21]). We compared Fig. 2b with Fig. 3b (upper left) and found that they were qualitatively similar. Fig. 3b (bottom right) shows the PMF curve between the solvophilic substrate and solvophilic probe. Unlike the PMF curve in Fig. 3b (upper left), there is a positive repulsive potential. Fig. 3b (upper right) shows the PMF curve between the solvophobic substrate and solvophilic probe. The shape of the PMF curve was qualitatively similar to that shown in Fig. 3b (bottom right). However, the repulsive potential in Fig. 3b (upper right) was lower than that in Fig. 3b (bottom right). Fig. 3b (bottom left) shows the PMF curve between the solvophilic substrate and solvophobic probe. The shape of the PMF curve is slightly similar to that shown in Fig. 3b (upper left). However, the attractive potential in Fig. 3b (bottom left) is shallower than that in Fig. 3b (upper left).

A comparison of Fig. 3b (upper right) with Fig. 3b (bottom left) shows that the shapes of both curves are definitely different from each other, despite mere exchange of the surface affinities of the substrate and probe. We call this behaviour the "asymmetric property". Here, we analogise *dispersion stability of the nanoparticles in liquid Ga* using Figs. 3a and 3b. For example, when the nanoparticles are solvophobic, their dispersion stability may be low because of attractive interactions. When the nanoparticles are solvophilic, their dispersion stability may be high owing to repulsive interactions. Interestingly, the solvophilic repulsive interaction arises at relatively long distances, the result of which cannot be obtained from thermodynamic theory[1]. This is an advantage of our statistical mechanics theory over the thermodynamic theory. Next, we considered two types of particles in the liquid Ga. One was a relatively large nanoparticle with a solvophobic surface, and the other was a relatively small nanoparticle with a solvophilic surface. From Figs. 3a and 3b (upper right), the repulsive interactions between the large and small nanoparticles can be analogised. Therefore, when the two types of particles are immersed in liquid Ga, there is a slight probability of forming aggregates consisting of solvophobic large nanoparticles and solvophilic small nanoparticles (Fig. 3c). However, we consider the following situation: one is a relatively large nanoparticle with a solvophilic surface and the other is a relatively small nanoparticle with a solvophobic surface. From Figs. 3a and 3b (bottom left), the attractive interactions between the large and small nanoparticles can be analogised. Hence, when these particles are immersed in liquid Ga, they may aggregate after a sufficient amount of time (Fig. 3d).

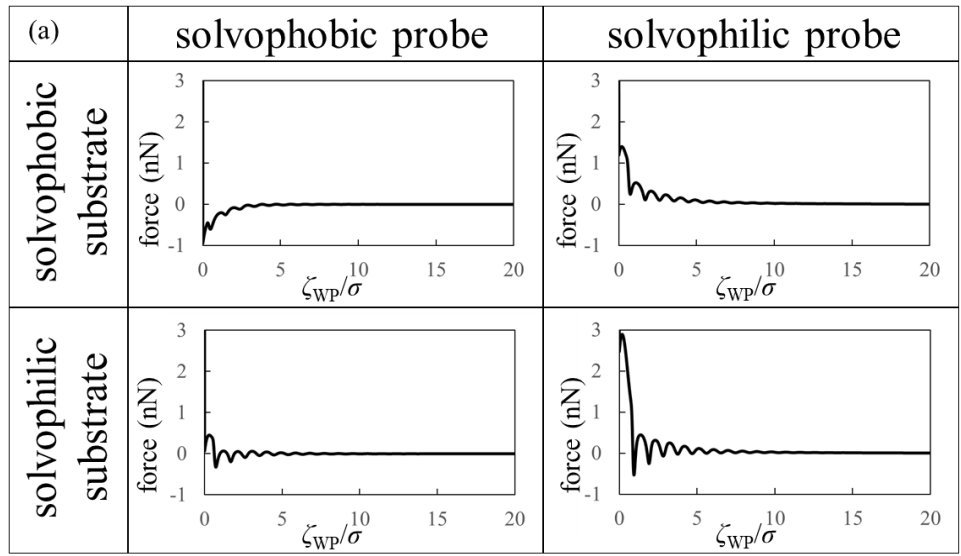

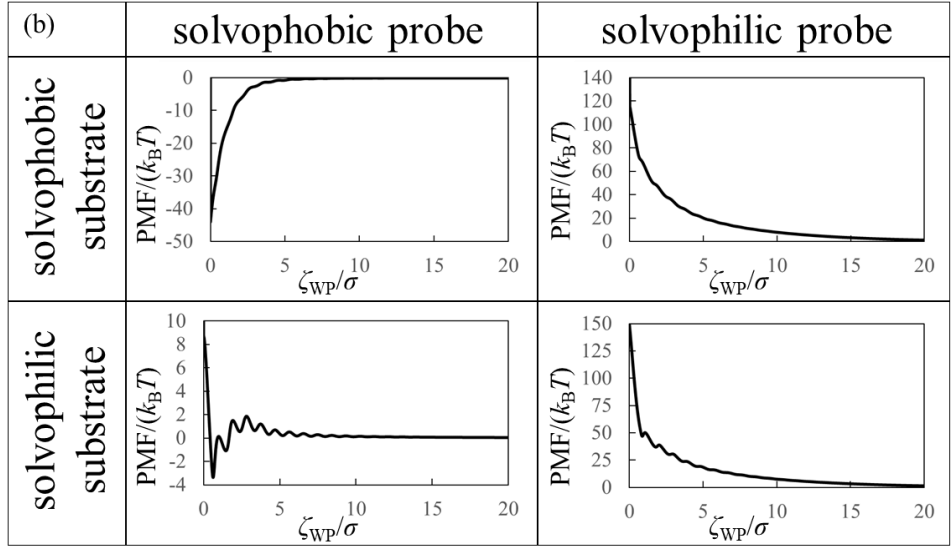

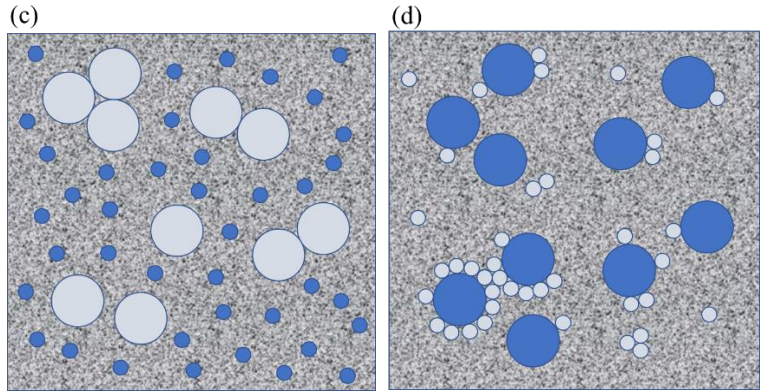

**Fig. 3| Theoretical results and dispersion/aggregation images. a**, Force curves, **b** PMF curves between the solvophobic/solvophilic substrate and solvophobic/solvophilic probe in liquid Ga. The diameter of the probe is five times the diameter of the Ga atom (*i.e.* $(2R + \sigma)/\sigma = 5$). $\zeta_{WP}$ is the distance between the closest surfaces of the substrate and probe (Fig. 1a). **c** and **d,** Schematics of the nanoparticles drawn by analogy with **a** and **b**. **c,** Schematic of the dispersed liquid containing large nanoparticles with solvophobic surfaces and small nanoparticles with solvophilic surfaces after adequate time. **d,** Schematic of the dispersed liquid containing large nanoparticles with solvophilic surfaces and small nanoparticles with solvophobic surfaces after adequate time.

In addition, we show the normalised number density distributions of the Ga atoms near the substrate ($g_{WGa}$) and probe ($g_{PGa}$) surfaces obtained from the OZ-HNC theory (Fig. 4). The solid and dashed curves show many oscillations. Moreover, their oscillation lengths are almost equal to the diameter of the Ga atom. Because there is a large difference in affinity, the heights of the first peaks exhibit a large difference. The shapes of both $g_{WGa}$ curves are qualitatively similar, but those of the force curves (Fig. 3a) and PMF curves (Fig. 3b) are apparently dissimilar. These trends are interesting because the force and PMF curves originate from $g_{WGa}$ curves with similar shapes.

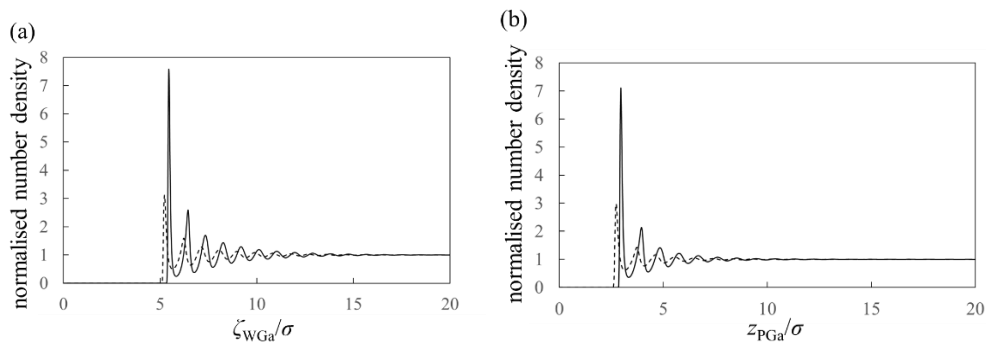

**Fig. 4| Solvation structures.** Normalised number densities of the liquid Ga near the **a,** substrate surface and **b,** probe surface. The solid and dashed curves are the normalised number densities on the surfaces of the strong and the weak affinities, respectively. $\zeta_{WGa}$ is the distance between the centres of the substrate and Ga atom (Fig. 1a). $z_{PGa}$ is the distance between the centres of the probe and Ga atom (Fig. 1a).

The experimental and theoretical force and PMF curves were obtained and compared, and the qualitative agreement between them was confirmed. Many oscillations and strong attractive/repulsive interactions were observed in the curves. From the theoretical calculations, we found an asymmetric property. We believe that these results are practical for controlling the dispersion and self-assembly of particles in liquid metals. This process will be helpful for the development of metals for artificial joints, reinforcing bars, and car bodies to add and enhance the desired properties. This will also be helpful for solving the problems of aggregation and precipitation of impurities in heat pipes. Furthermore, it is expected that the surfaces of nanoparticles dispersed in a liquid metal can be used as adsorption layers for some substances. In the future, we intend to develop an inverse analysis theory for liquid metal to clarify the solvation structure on the substrate from the experimentally obtained AFM force curve. Moreover, we plan to inversely calculate[16,22] the pair potential between the substrate and atoms of the liquid metal to discover the existence of Friedel oscillations between them. Inverse calculations are considered to be an important first step toward the elucidation of the Friedel oscillation between them.

**Methods**

**AFM experiment**

The AFM experiments were performed under atmospheric conditions at room temperature using a frequency modulation (FM) technique, in which the forces acting on the sensor were detected as a resonance frequency shift ($\Delta f$)[23]. A commercial AFM (JEOL Co.; JSPM-5200) with a Nanonis AFM control system (SPECS Zurich GmbH) was used, with some modifications. The AFM force sensor used in this study was a qPlus sensor fabricated using a commercial quartz tuning fork (STATEK Co. TFW-1165) and an electrochemically etched tungsten probe[24,25]. The resonance frequency and spring constant of the tuning fork were 32.768 kHz and 1884 N m$^{-1}$, respectively. The probe was made of a tungsten wire with diameter of 0.1 mm (Nilaco Co.) and was etched in a potassium hydroxide solution of 1.2 mol L$^{-1}$. The qPlus sensor was mechanically vibrated by a lead zirconate titanate (PZT) piezoelectric actuator, and its deflection was electrically detected by a differential current amplifier embedded in the AFM head[26]. The $\mathit{\Delta f}$ of the qPlus sensor was detected using a commercial FM demodulator (Kyoto Instruments, KI-2001) with some modifications[21], where the vibrating amplitude was kept constant by home-built feedback electronics.

Liquid Ga droplet (several microlitres) was deposited on a cleaved mica surface under atmospheric conditions[13]. Only the probe apex was inserted into the liquid Ga droplet. The Ga surface was oxidised instantaneously in the atmosphere. The oxide film formed on the Ga/air interface penetrated the probe, and the interface between the mica surface and liquid Ga was investigated using AFM.

An epitaxially-grown Au(111) film with a 100 nm thickness was prepared by vacuum evaporation (base pressure: ~10$^{-5}$ Pa, evaporation rate: ~0.1 nm sec$^{-1}$) on a cleaved mica substrate. During the evaporation, the substrate temperature was kept at 723K. A several μL of liquid Ga droplet was deposited on a cleaved mica or the Au film under an atmospheric condition[13]. The liquid Ga started to diffuse into the Au film shortly after the deposition, and Au-Ga alloy was formed. The AFM investigation on the Au-Ga alloy was carried out approximately 10 hours after the deposition of the liquid Ga, which was enough for the formation of the AuGa$_2$ alloy crystals[13]. Then, the interface between the alloy surface and the liquid Ga was investigated using the AFM.

The $\mathit{\Delta f}$-distance curve was obtained by changing the tip-to-sample distance without the $\mathit{\Delta f}$ feedback at a constant surface position, and it was converted to the force-distance curve using the method developed by Sader and Jarvis[27].

However, two aspects must be considered. First, it is probable that AFM did not analyze the mica surface, but rather the Ga oxide film that covered the mica substrate. As described above, liquid Ga was dropped onto a mica surface in air. Therefore, the surface of the Ga droplet was already covered with the Ga oxide film, and it is likely that the oxide film covered the mica surface when the droplet was placed. Similarly, it is possible that the probe was covered with a Ga oxide film. The second is the contact angle of the Ga droplets, which could not be measured accurately in air because of the oxide film growth. As the oxide film continued to grow under atmospheric conditions, the contact angle of the Ga droplet changed. Therefore, the contact angle of the Ga droplet measured in air was different from the true contact angle. However, when the Ga droplet was placed on the mica substrate, the wettability of the droplet was poor. Furthermore, the wettability of liquid Ga is very poor for many non-metallic materials such as quartz, sapphire, and graphite[28]. Moreover, although it is a property of Galinstan (a mixture of Ga, In, and Sn), the contact angles of the mixed liquid on metallic and nonmetallic materials are reported to be larger than 90°[28]. Therefore, it is reasonable to assume that the true contact angles of the liquid Ga on the surfaces of the mica substrate, tungsten probe, and oxide films coated on the substrate and probe were greater than 90°.

**Pair potentials**

To understand the mechanism of interaction between the substrate and probe in liquid Ga, we conducted computational calculations. For the calculation, we prepared a simple system (Fig. 1a) and focused on the following four pair potentials. The pair potentials are shown in the *Supplementary Information*.

The pair potential between the atoms of the liquid Ga is explained below. Liquid Ga contains Ga cations and conduction electrons. It is difficult to treat the conduction electrons in liquid Ga explicitly theoretically. Hence, we used the effective potential between the Ga cations in the sea of conduction electrons provided by Waseda *et al*[8] (Fig. 1b). The effect of conduction electrons is implicitly included in the pair potential. As shown in Fig. 1b, there are many oscillations caused by the Friedel oscillation[29]. The reason for selecting the Waseda pair potential is as follows. Several pair potentials exist between the metal atoms[6730313233]. These pair potentials have been theoretically proposed, whereas those of Waseda *et al.* have been experimentally determined; it considers the real liquid Ga condition. Hence, we selected the Waseda pair potential.

The pair potential between the substrate and the atom of liquid Ga is explained below. The pair potential between the substrate and Ga atom $U_{WGa}$ was modelled using the following equation[3435]:

$$U_{WGa}(z_{WGa}) = 2\pi\varepsilon_{WGa}\left[\frac{2}{5}\left(\frac{\sigma}{z_{WGa}}\right)^{10} - \left(\frac{\sigma}{z_{WGa}}\right)^4 - \frac{\sigma^4}{3\Delta(z_{WGa} + 0.61\Delta)^3}\right], \quad (1)$$

where $z_{WGa}$, $\sigma$, and $\Delta$ are the distances between the centres of the substrate surface atom and Ga atom, the diameter of the Ga atom (0.255 nm)[8], and $\sigma/\sqrt{2}$, respectively. Parameter $\varepsilon_{WGa}$ represents the affinity between the substrate and the Ga atom, where we prepared both the weak (solvophobic) and strong (solvophilic) parameters ($\varepsilon_{WGa} = 10^{-22}$ or $75 \times 10^{-22}$ J). The pair potential between the substrate and Ga atom is shown in Fig. S3a.

The pair potential between the probe and liquid Ga atom is explained below. The pair potential between the probe and Ga atom, which is $U_{PGa}$, was modelled using the following equation[3435]:

$$U_{\text{PGa}}(z_{\text{PGa}}) = 2\pi\varepsilon_{\text{PGa}}\left[\frac{2}{5}\left(\frac{\sigma}{z_{\text{PGa}} - R}\right)^{10} - \left(\frac{\sigma}{z_{\text{PGa}} - R}\right)^{4} - \frac{\sigma^{4}}{3\Delta(z_{\text{PGa}} - R + 0.61\Delta)^{3}}\right], \quad (2)$$

where $z_{\text{PGa}}$ is the distance between the centres of the probe and Ga atom. Parameter $R$ is the distance between the centres of the probe and probe surface atom. The parameter $\varepsilon_{\text{PGa}}$ represents the affinity between the probe and Ga atom. Again, we prepared both weak (solvophobic) and strong (solvophilic) parameters ($\varepsilon_{\text{PGa}} = 10^{-22}$ or $75 \times 10^{-22}$ J). The pair potential between the probe and Ga atom is shown in Figs. S3b and Sc.

The pair potential between the substrate and probe is explained below. The pair potential between the substrate and probe, which is $U_{\text{WP}}$, was modelled as follows:[34,36]

$$U_{\text{WP}}(z_{\text{WP}}) = \begin{cases} \infty & (z_{\text{WP}} < \sigma), \\ -\dfrac{A_{\text{WP}}}{6}\left\{\dfrac{R}{z_{\text{WP}}} + \dfrac{R}{z_{\text{WP}} + 2R} + \ln\left(\dfrac{z_{\text{WP}}}{z_{\text{WP}} + 2R}\right)\right\} & (z_{\text{WP}} \geq \sigma), \end{cases} \quad (3)$$

where $A_{\text{WP}}$ and $z_{\text{WP}}$ are the Hamaker constant between the substrate and probe in the continuum liquid Ga and the distance between the centres of the substrate surface atom and probe surface atom, respectively. The pair potential between the substrate and probe is shown in Fig. S3d. From the AFM experiment, we considered that the surfaces of the substrate and the probe were composed of the gallium oxide; hence, we set $A_{\text{WP}} = 8.174 \times 10^{-19}$ J. The value of $A_{\text{WP}}$ was calculated using the following equation[36]:

$$A_{\text{WP}} = \left(\sqrt{A_{\text{Ga}_2\text{O}_3}} - \sqrt{A_{\text{Ga}}}\right)^{2}, \quad (4)$$

where $A_{\text{Ga}_2\text{O}_3}$ and $A_{\text{Ga}}$ are the Hamaker constants of solid $Ga_2O_3$ and liquid Ga in vacuum,

respectively. The Hamaker constant ($A_{Ga_2O_3}$ or $A_{Ga}$) can be expressed using the corresponding surface tension, as follows:

$$A = 24\pi l^2 \gamma, \tag{5}$$

where $A$, $l$, and $\gamma$ are the Hamaker constant of the substance of interest, the length of one side of a cube containing one molecule, the diameter of the molecule of interest, and the surface tension between solid $Ga_2O_3$ and vacuum (inert gas) or liquid Ga and vacuum (inert gas), respectively. Yunusa et al.[37] reported the *apparent* surface tension between $Ga_2O_3$ and the gas as 591 mN/m; hence, we applied the value for the surface tension. The apparent surface tension is not exactly the same as that between solid $Ga_2O_3$ and gas; however, we used this value because when liquid $Ga_2O_3$ is placed on solid $Ga_2O_3$, the contact angle should be close to *zero*. In this case, $\gamma_{sg}$ (surface tension between the solid and gas) is nearly equal to $\gamma_{lg}$ (surface tension between the liquid and gas). Therefore, a value of 591 mN/m was applied for $\gamma_{sg}$. The surface tension between liquid Ga and gas was 695 mN/m[37]. Next, we explain $l$ in Eq. (5) to obtain its value. The number density of amorphous $Ga_2O_3$ is $1.3 \times 10^{28}$ m$^{-3}$[19]. (The number density of $\beta$-$Ga_2O_3$ has been reported[19] to be $1.9 \times 10^{28}$ m$^{-3}$. However, the oxide film on liquid Ga is amorphous or poorly crystallised[18]. Hence, we used the value of $1.3 \times 10^{28}$ m$^{-3}$. Although not shown, the conclusion of the present study does not change even when the value is $1.9 \times 10^{28}$ m$^{-3}$.) From the number density, the value of $l$ for amorphous $Ga_2O_3$ was estimated to be 0.425 nm. In our computational calculation of the interactions between the substrate and the probe in liquid Ga, the temperature was set at 60 °C to converge the computation. Hence, we used the following number density of liquid Ga at 60 °C: $5.245 \times 10^{28}$ m$^{-3}$[38]. Accordingly, the value of $l$ for liquid Ga is 0.267 nm.

**Integral equation theory**

We employed integral equation theory (statistical mechanics of simple liquids) and applied the Ornstein–Zernike (OZ) equation coupled with a hypernetted-chain (HNC) closure named OZ-HNC[14,15,16,17]. Using the OZ-HNC theory, the normalised number density distributions of liquid Ga near the substrate and the interactions (force and PMF) between the substrate probe in the liquid Ga were calculated. The calculation temperature was 60 °C to avoid crystallisation during calculation. Because our calculation incorporates the Waseda pair potential, this calculation method implicitly considers the effects of conduction electrons in the liquid metal.

## References

(1) Chen, L. Y.; Xu, J. Q.; Choi, H.; Pozuelo, M.; Ma, X.; Bhowmick, S.; Yang, J. M.; Mathaudhu, S.; Li, X. C. Processing and Properties of Magnesium Containing a Dense Uniform Dispersion of Nanoparticles. *Nature* **2015**, *528* (7583), 539–543. https://doi.org/10.1038/nature16445.

(2) Zhang, S.; Liu, Y.; Fan, Q.; Zhang, C.; Zhou, T.; Kalantar-Zadeh, K.; Guo, Z. Liquid Metal Batteries for Future Energy Storage. *Energy Environ. Sci.* **2021**, *14* (8), 4177–4202. https://doi.org/10.1039/d1ee00531f.

(3) Yun, G.; Tang, S. Y.; Zhao, Q.; Zhang, Y.; Lu, H.; Yuan, D.; Sun, S.; Deng, L.; Dickey, M. D.; Li, W. Liquid Metal Composites with Anisotropic and Unconventional Piezoconductivity. *Matter* **2020**, *3* (3), 824–841. https://doi.org/10.1016/j.matt.2020.05.022.

(4)     Faghri, A. Heat Pipes: Review, Opportunities and Challenges. *Front. Heat Pipes* **2014**, *5* (1). https://doi.org/10.5098/fhp.5.1.

(5)     El-Genk, M. S.; Tournier, J.-M. P. Uses of Liquid-Metal and Water Heat Pipes in Space Reactor Power Systems. *Front. Heat Pipes* **2011**, *2* (1). https://doi.org/10.5098/fhp.v2.1.3002.

(6)     Plimpton, S. J.; Wolf, E. D. Effect of Interatomic Potential on Simulated Grain-Boundary and Bulk Diffusion: A Molecular-Dynamics Study. *Phys. Rev. B* **1990**, *41* (5), 2712–2721. https://doi.org/10.1103/PhysRevB.41.2712.

(7)     Zhao, M.; Chekmarev, D. S.; Cai, Z. H.; Rice, S. A. Structure of Liquid Ga and the Liquid-Vapor Interface of Ga. *Phys. Rev. E - Stat. Physics, Plasmas, Fluids, Relat. Interdiscip. Top.* **1997**, *56* (6), 7033–7042. https://doi.org/10.1103/PhysRevE.56.7033.

(8)     Waseda, Y.; Ohtani, M. Effective Interionic Potentials and Properties of Molten Metals. *J. Japan Inst. Met. Mater.* **1972**, 1016–1025.

(9)     Ascarelli, P. Atomic Radial Distributions and Ion-Ion Potential in Liquid Gallium. *Phys. Rev.* **1966**, *143* (1), 36–47. https://doi.org/10.1103/PhysRev.143.36.

(10)    Edwards, D. J.; Jarzynski, J. Ion-Ion Potentials in Liquid Metals. *J. Phys. C Solid State Phys.* **1972**, *5* (14), 1745–1756. https://doi.org/10.1088/0022-3719/5/14/004.

(11)    Liu, Z.; Ma, M.; Liang, W.; Deng, H. A Mechanistic Study of Clustering and Diffusion of Molybdenum and Rhenium Atoms in Liquid Sodium. *Metals (Basel).* **2021**, *11* (9), 1–11. https://doi.org/10.3390/met11091430.

(12)    Wang, J.; Horsfield, A.; Schwingenschlögl, U.; Lee, P. D. Heterogeneous Nucleation of Solid Al from the Melt by TiB2 and Al3 Ti: An Ab Initio Molecular Dynamics Study. *Phys. Rev. B - Condens. Matter Mater. Phys.* **2010**, *82* (18), 1–10. https://doi.org/10.1103/PhysRevB.82.184203.

(13) Ichii, T.; Murata, M.; Utsunomiya, T.; Sugimura, H. Atomic-Scale Structural Analysis on the Interfaces between Molten Gallium and Solid Alloys by Atomic Force Microscopy. *J. Phys. Chem. C* **2021**, *125* (47), 26201–26207. https://doi.org/10.1021/acs.jpcc.1c08029.

(14) Amano, K. I.; Liang, Y.; Miyazawa, K.; Kobayashi, K.; Hashimoto, K.; Fukami, K.; Nishi, N.; Sakka, T.; Onishi, H.; Fukuma, T. Number Density Distribution of Solvent Molecules on a Substrate: A Transform Theory for Atomic Force Microscopy. *Phys. Chem. Chem. Phys.* **2016**, *18* (23), 15534–15544. https://doi.org/10.1039/c6cp00769d.

(15) Amano, K. I.; Ishihara, T.; Hashimoto, K.; Ishida, N.; Fukami, K.; Nishi, N.; Sakka, T. Stratification of Colloidal Particles on a Surface: Study by a Colloidal Probe Atomic Force Microscopy Combined with a Transform Theory. *J. Phys. Chem. B* **2018**, *122* (16), 4592–4599. https://doi.org/10.1021/acs.jpcb.8b01082.

(16) Hashimoto, K.; Amano, K. ichi; Nishi, N.; Sakka, T. Integral Equation Theory Based Method to Determine Number Density Distribution of Colloidal Particles near a Substrate Using a Force Curve from Colloidal Probe Atomic Force Microscopy. *J. Mol. Liq.* **2019**, *294*, 111584. https://doi.org/10.1016/j.molliq.2019.111584.

(17) Furukawa, S.; Amano, K. ichi; Ishihara, T.; Hashimoto, K.; Nishi, N.; Onishi, H.; Sakka, T. Enhancement of Stratification of Colloidal Particles near a Substrate Induced by Addition of Non-Adsorbing Polymers. *Chem. Phys. Lett.* **2019**, *734*, 1–9.

(18) Regan, M. J.; Tostmann, H.; Pershan, P. S.; Magnussen, O. M.; Dimasi, E.; Ocko, B. M.; Deutsch, M. X-Ray Study of the Oxidation of Liquid-Gallium Surfaces. **1997**, *55* (16), 786–790.

(19) Kobayashi, E.; Boccard, M.; Jeangros, Q.; Rodkey, N.; Vresilovic, D.; Hessler-Wyser, A.; Döbeli, M.; Franta, D.; De Wolf, S.; Morales-Masis, M.; Ballif, C. Amorphous

Gallium Oxide Grown by Low-Temperature PECVD. *J. Vac. Sci. Technol. A Vacuum, Surfaces, Film.* **2018**, *36* (2), 021518. https://doi.org/10.1116/1.5018800.

(20) Kaggwa, G. B.; Nalam, P. C.; Kilpatrick, J. I.; Spencer, N. D.; Jarvis, S. P. Impact of Hydrophilic/Hydrophobic Surface Chemistry on Hydration Forces in the Absence of Confinement. *Langmuir* **2012**, *28* (16), 6589–6594. https://doi.org/10.1021/la300155c.

(21) Kobayashi, K.; Yamada, H.; Itoh, H.; Horiuchi, T.; Matsushige, K. Analog Frequency Modulation Detector for Dynamic Force Microscopy. *Rev. Sci. Instrum.* **2001**, *72* (12), 4383–4387. https://doi.org/10.1063/1.1416104.

(22) Hashimoto, K.; Amano, K. ichi; Nishi, N.; Sakka, T. Calculation Method of the Number Density Distribution of Liquid Molecules or Colloidal Particles near a Substrate from Surface Force Apparatus Measurement. *Chem. Phys. Lett.* **2020**, *754* (June), 137666. https://doi.org/10.1016/j.cplett.2020.137666.

(23) Albrecht, T. R.; Grütter, P.; Horne, D.; Rugar, D. Frequency Modulation Detection Using High-Q Cantilevers for Enhanced Force Microscope Sensitivity. *J. Appl. Phys.* **1991**, *69* (2), 668–673. https://doi.org/10.1063/1.347347.

(24) Giessibl, F. J. High-Speed Force Sensor for Force Microscopy and Profilometry Utilizing a Quartz Tuning Fork. *Appl. Phys. Lett.* **1998**, *73* (26), 3956–3958. https://doi.org/10.1063/1.122948.

(25) Giessibl, F. J. Atomic Resolution on Si(111)-(7×7) by Noncontact Atomic Force Microscopy with a Force Sensor Based on a Quartz Tuning Fork. *Appl. Phys. Lett.* **2000**, *76* (11), 1470–1472. https://doi.org/10.1063/1.126067.

(26) Huber, F.; Giessibl, F. J. Low Noise Current Preamplifier for QPlus Sensor Deflection Signal Detection in Atomic Force Microscopy at Room and Low Temperatures. *Rev. Sci. Instrum.* **2017**, *88* (7). https://doi.org/10.1063/1.4993737.

(27) Sader, J. E.; Jarvis, S. P. Accurate Formulas for Interaction Force and Energy in Frequency Modulation Force Spectroscopy. *Appl. Phys. Lett.* **2004**, *84* (10), 1801–1803. https://doi.org/10.1063/1.1667267.

(28) Naidich, J. V.; Chuvashov, J. N. Wettability and Contact Interaction of Gallium-Containing Melts with Non-Metallic Solids. *J. Mater. Sci.* **1983**, *18* (7), 2071–2080. https://doi.org/10.1007/BF00555000.

(29) Overhauser, A. W. Surface Friedel Oscillations and Photoemission from Simple Metals. *Phys. Rev. B* **1986**, *33* (2), 1468–1470.

(30) Hafner, J. Structure of Liquid Arsenic: Peierls Distortion versus Friedel Modulation. *Phys. Rev. Lett.* **1989**, *62* (7), 784–787.

(31) Sumi, T.; Miyoshi, E.; Tanaka, K. Molecular-Dynamics Study of Liquid Mercury in the Density Region between Metal and Nonmetal. *Phys. Rev. B - Condens. Matter Mater. Phys.* **1999**, *59* (9), 6153–6158. https://doi.org/10.1103/PhysRevB.59.6153.

(32) Belashchenko, D. K.; Ostrovskii, O. I. The Embedded Atom Model for Liquid Metals: Liquid Gallium and Bismuth. *Russ. J. Phys. Chem. A* **2006**, *80* (4), 509–522. https://doi.org/10.1134/S0036024406040054.

(33) Ghatee, M. H.; Karimi, H.; Shekoohi, K. Structural, Mechanical and Thermodynamical Properties of Silver Amalgam Filler: A Monte Carlo Simulation Study. *J. Mol. Liq.* **2015**, *211*, 96–104. https://doi.org/10.1016/j.molliq.2015.06.062.

(34) Pacheco, J. M.; Ekardt, W. Microscopic Calculation of the van Der Waals Interaction between Small Metal Clusters. *Phys. Rev. Lett.* **1992**, *68* (25), 3694–3697.

(35) Shinto, H.; Uranishi, K.; Miyahara, M.; Higashitani, K. Wetting-Induced Interaction between Rigid Nanoparticle and Plate: A Monte Carlo Study. *J. Chem. Phys.* **2002**, *116* (21), 9500–9509. https://doi.org/10.1063/1.1473817.

(36) Israelachvili, J. N. *Intermolecular and Surface Forces*, Third edit.; Academic Press, 2011.

(37) Yunusa, M.; Amador, G. J.; Drotlef, D. M.; Sitti, M. Wrinkling Instability and Adhesion of a Highly Bendable Gallium Oxide Nanofilm Encapsulating a Liquid-Gallium Droplet. *Nano Lett.* **2018**, *18* (4), 2498–2504. https://doi.org/10.1021/acs.nanolett.8b00164.

(38) Okada, K.; Ozoe, H. Transient Responses of Natural Convection Heat Transfer with Liquid Gallium under an External Magnetic Field in Either the y, or z Direction. *Ind. Eng. Chem. Res.* **1992**, *31* (3), 700–706. https://doi.org/10.1021/ie00003a008.

## Data availability

The data supporting the findings of this study are available from the corresponding authors upon reasonable request.

## Acknowledgements

This work was supported by KAKENHI (20H02619), and partially supported by KAKENHI (20K05437). We thank M. Maebayashi for providing a computational research environment.

## Author contributions

All of the authors contributed to preparation of the manuscript. T.I. proposed the experimental direction of this research and K.A. and H.N. proposed the theoretical direction of this research. I.T. developed the FM-AFM. I.T., M.M., and Y.A. measured the force curve. K.A. developed the calculation program. K.A., K.T. and M.T. performed the computation and prepared the figures. T.I., T.U., and H.S. investigated the experimental background. K.A., K.T., M.T., and H.N. investigated the theoretical background.

## Competing interests

The authors declare no competing interests.

## Corresponding authors

Correspondence to Ken-ichi Amano (amanok@meijo-u.ac.jp) or Tkashi Ichii (ichii.takashi.2m@kyoto-u.ac.jp).

# Supplementary information

# Interaction between substrate and probe in liquid metal Ga: Experimental and theoretical analysis

**Authors:** Ken-ichi Amano[1,✉], Kentaro Tozawa[1], Maho Tomita[1], Hiroshi Nakano[2], Makoto Murata[3], Yousuke Abe[3], Toru Utsunomiya[3], Hiroyuki Sugimura[3], and Takashi Ichii[3,✉]

[1]Faculty of Agriculture, Meijo University, Nagoya, Aichi 468-8502, Japan

[2]Department of Molecular Engineering, Kyoto University, Kyoto 615-8510, Japan

[3]Department of Materials Science and Engineering, Kyoto University, Kyoto 606-8501, Japan

✉e-mail: amanok@meijo-u.ac.jp; ichii.takashi.2m@kyoto-u.ac.jp

In the supplementary information, we show the experimental and the theoretical data in Figs. 1,2 and Figs. 3-5, respectively. For the experimental detail, see Methods subsection 'AFM experiment'. For the theoretical detail, see Methods subsections 'Pair potentials' and 'Integral equation theory'.

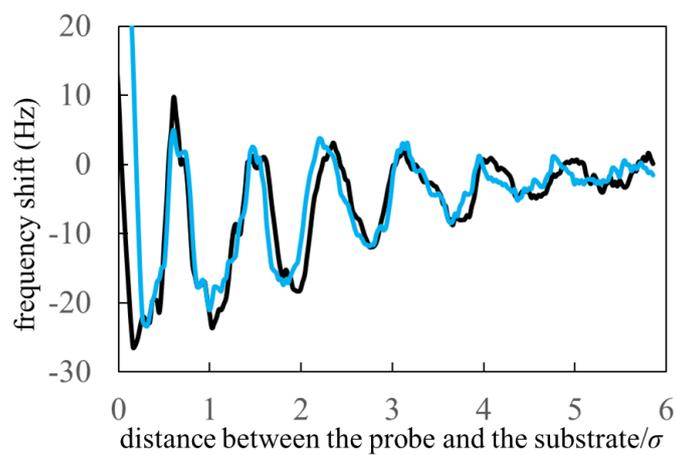

**Fig. S1**. Frequency shift curves measured by our AFM. Substrate (mica) and probe (tungsten) are covered with gallium oxide film[1–3]. Black and blue curves represent the approach and retract curves.

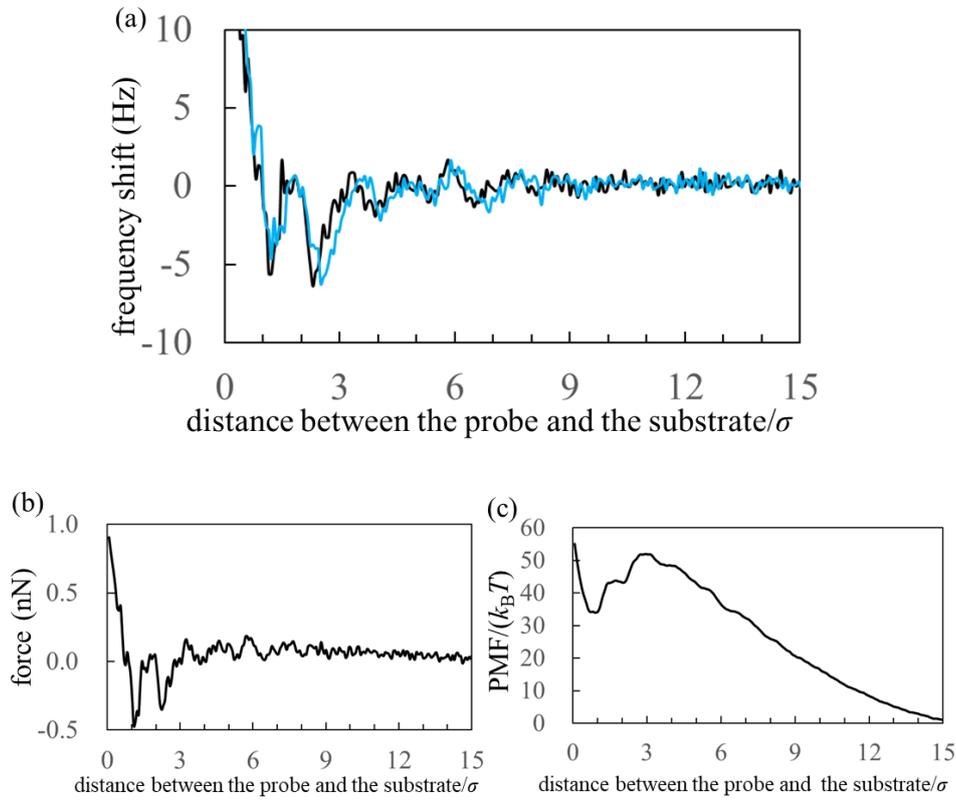

**Fig. S2**. (a) Frequency shift, (b) force, (c) PMF curves measured by our AFM on AuGa$_2$ surface. Probe is tungsten which is covered with gallium oxide film. Black and blue represent the approach and retract curves. The experimental condition may correspond to the solvophilic substrate and solvophobic probe pair. In (c), increase in PMF is observed from 15$\sigma$ to 3$\sigma$, while decrease in PMF is observed from 3$\sigma$ to 1$\sigma$. This behaviour is very similar to those in Fig. 3b (bottom left) and Fig. S4b (bottom left). However, the oscillation lengths in (b) and (c) is about 1.5 times larger than $\sigma$. This is because the AuGa2 crystal surface plane was tilted with respect to the scanning direction of the probe[3]. Although it is difficult to control the direction of the crystal surface, we consider that the oscillation length approaches to 1$\sigma$ when the crystal surface and the scanning direction is vertical.

The paired potentials between the substrate and Ga atom, the probe and Ga atom, and the substrate and probe are expressed by Eqs. (1), (2), and (3), as shown in Figs. S3(a), S3(b)(c), and S3(d), respectively.

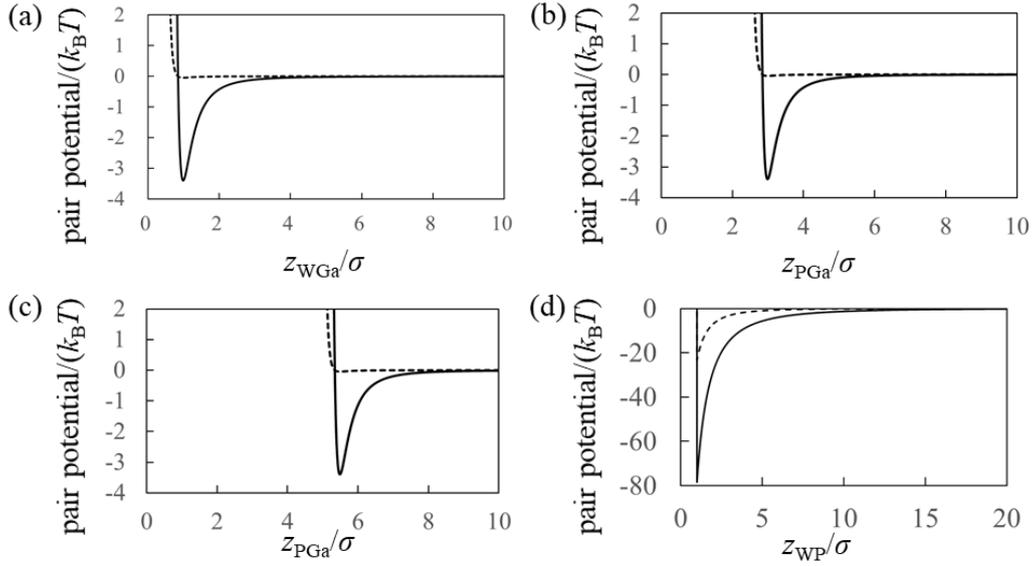

**Fig. S3**. (a) Solid and dashed curves are the pair potentials between the substrate and Ga atom when the affinity parameters are strong and weak, respectively. (b), (c) Solid and dashed curves represent the pair potentials between the probe and Ga atom when the affinity parameters are strong and weak. In (b), the diameter of the probe is five times that of the Ga atom (i.e. $(2R + \sigma)/\sigma = 5$). In (c), the diameter of the probe is 10 times that of the Ga atom (i.e. $(2R + \sigma)/\sigma = 10$). (d) The solid and dashed curves denote the pair potentials between the substrate and probe when the probe sizes are 10 and 5 times that of the liquid Ga atom, respectively. $k_B$ and $T$ are the Boltzmann constant and absolute temperature, respectively, where $T$ is 333.15 K (i.e. 60°C).

Here, the diameter of the probe was 10 times the diameter of the Ga atom (*i.e.* $(2R + \sigma)/\sigma = 10$). In this case, the force curves, PMF curves, and density distributions of the Ga atoms near the probe calculated from the OZ-HNC theory are shown in Figs. S4 and S5. Affinity parameters $\varepsilon_{WGa}$ (= $10^{-22}$ or $75 \times 10^{-22}$ J) and $\varepsilon_{PGa}$ (= $10^{-22}$ or $75 \times 10^{-22}$ J) were used. We define values of the affinity parameters $10^{-22}$ J and $5 \times 10^{-21}$ J as "solvophobic" and "solvophilic", respectively.

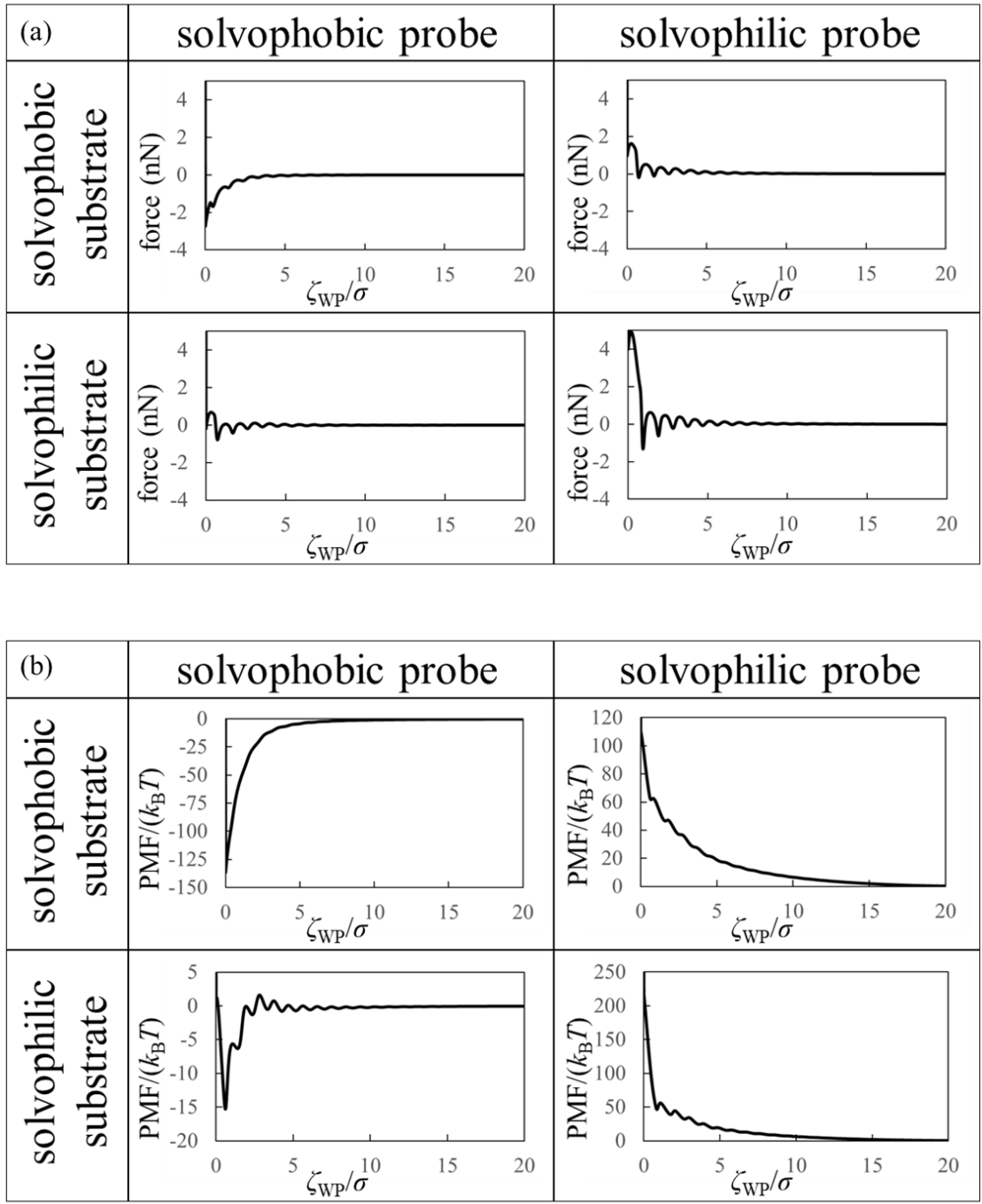

**Fig. S4**. (a) Force curves and (b) PMF curves between the solvophobic/solvophilic substrate and the solvophobic/solvophilic probe in the liquid Ga when the probe size is 10 times that of the liquid Ga atom. $\zeta_{WP}$ is the distance between the closest surfaces of the substrate and the probe (see Fig. 1a).

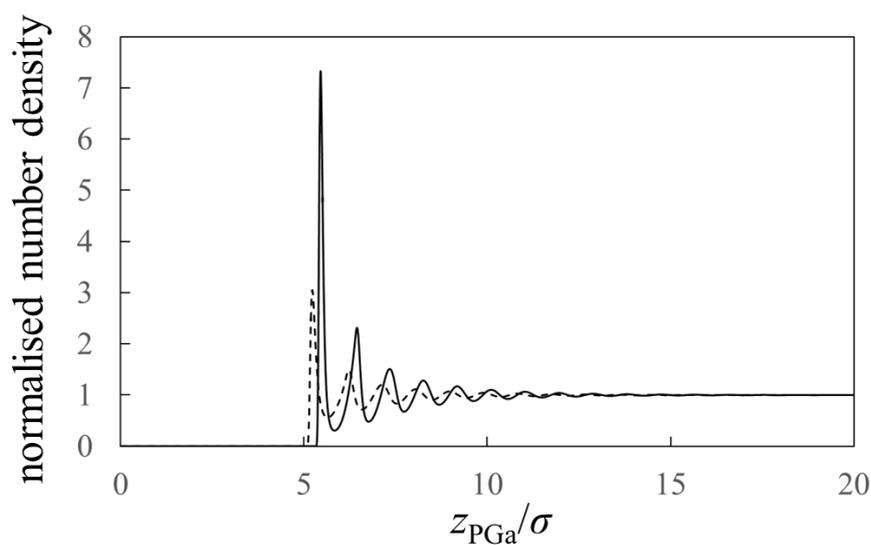

**Fig. S5**. Normalised number densities of the liquid Ga near the probe surface when the probe size is 10 times that of the liquid Ga atom. The solid and dashed curves denote the normalised number densities on the solvophilic and solvophobic surfaces, respectively. $z_{PGa}$ is the distance between the centres of the probe and the Ga atom (see Fig. 1a).

# References

1. Regan, M. J. *et al.* X-ray study of the oxidation of liquid-gallium surfaces. **55**, 786–790 (1997).

2. Kobayashi, E. *et al.* Amorphous gallium oxide grown by low-temperature PECVD. *J. Vac. Sci. Technol. A Vacuum, Surfaces, Film.* **36**, 021518 (2018).

3. Ichii, T., Murata, M., Utsunomiya, T. & Sugimura, H. Atomic-Scale Structural Analysis on the Interfaces between Molten Gallium and Solid Alloys by Atomic Force Microscopy. *J. Phys. Chem. C* **125**, 26201–26207 (2021).